\documentclass[sigconf,anonymous=false,review=false,natbib=true,authorversion=True]{acmart}

\usepackage[show]{chato-notes}
\usepackage{wrapfig}
\usepackage{multirow}
\usepackage{multicol}
\usepackage{makecell}
\usepackage{enumitem}
\usepackage{graphicx}
\usepackage{placeins}
\usepackage{microtype}
\usepackage{subfig}
\usepackage[para]{footmisc}
\usepackage{colortbl}
\usepackage{mathtools}

\usepackage{hyperref}
\usepackage{nameref}

\newcounter{mylabelcounter}

\makeatletter
\newcommand{\labeltext}[2]{%
    #1\refstepcounter{mylabelcounter}%
    \immediate\write\@auxout{%
        \string\newlabel{#2}{{\unexpanded{#1}}{\thepage}{{\unexpanded{#1}}}{mylabelcounter.\number\value{mylabelcounter}}{}}%
    }
}
\makeatother

\graphicspath{ {./graphics/} }

\usepackage{marginnote}
\newcommand{\pageenlarge}[1]{\marginnote{}\enlargethispage{#1\baselineskip}}

\newcommand{\craig}[1]{\textcolor{black}{#1}}
\newcommand{\sff}[1]{\textcolor[HTML]{000000}{#1}}
\newcommand{\sn}[1]{\textcolor[HTML]{000000}{#1}}
\newcommand{\sdt}[1]{\textcolor[HTML]{000000}{#1}}
\newcommand{\swsdm}[1]{\textcolor[HTML]{000000}{#1}}
\newcommand{\sd}[1]{\textcolor{black}{#1}}

\newcommand{\crc}[1]{\textcolor{black}{#1}}
\newcommand{\crcb}[1]{\textcolor{black}{#1}}
\newcommand{\crcf}[1]{\textcolor{black}{#1}}
\newcommand{\crcs}[1]{\textcolor{black}{#1}}

\title{RecJPQ: Training Large-Catalogue Sequential Recommenders}

\author{Aleksandr V. Petrov}
\affiliation{%
  \institution{University of Glasgow} \country{United Kingdom}}

\email{a.petrov.1@research.gla.ac.uk}

\author{Craig Macdonald}
\affiliation{%
  \institution{University of Glasgow} \country{United Kingdom}}
\email{craig.macdonald@glasgow.ac.uk}

\copyrightyear{2024}
\acmYear{2024}
\setcopyright{acmlicensed}\acmConference[WSDM '24]{Proceedings of the 17th ACM International Conference on Web Search and Data Mining}{March 4--8, 2024}{Merida, Mexico}
\acmBooktitle{Proceedings of the 17th ACM International Conference on Web Search and Data Mining (WSDM '24), March 4--8, 2024, Merida, Mexico}
\acmPrice{15.00}
\acmDOI{10.1145/3616855.3635821}
\acmISBN{979-8-4007-0371-3/24/03}

\begin{document}
\begin{abstract}
\looseness -1 \crcf{Sequential Recommendation is a popular recommendation task that uses the order of user-item interaction to model evolving users' interests and sequential patterns in their behaviour. Current state-of-the-art  Transformer-based models} \crcb{for sequential recommendation,} such as BERT4Rec and SASRec\crcb{,} generate sequence embeddings and compute scores for catalogue items, but the increasing catalogue size makes training these models costly. The Joint Product Quantisation (JPQ) method, \swsdm{originally proposed for passage retrieval}, markedly reduces the size of the retrieval index with minimal effect on model effectiveness\crcb{,} by replacing passage embeddings with a limited number of shared \crcb{sub-embeddings}. This paper introduces RecJPQ, a novel adaptation of JPQ for sequential recommendations\crcb{, which takes the place of item embeddings tensor and replaces item embeddings with a concatenation of a  limited number of shared sub-embeddings and, therefore, limits the number of learnable model parameters. The main idea of RecJPQ is to split items into sub-item entities before training the main recommendation model, which is inspired by splitting words into tokens and training tokenisers in language models}. 
We apply RecJPQ to SASRec, BERT4Rec, and GRU4rec models on three large-scale sequential datasets. Our results showed that RecJPQ \crcb{could} notably reduce \crcb{the} model size (e.g., 48$\times$ reduction for the Gowalla dataset with no effectiveness degradation). RecJPQ can also improve model performance through a regularisation effect (e.g.\ +0.96\% NDCG@10 improvement on the Booking.com dataset). \crcs{Overall, RecJPQ \crcs{allows the training of state-of-the-art transformer recommenders in industrial applications, where datasets with millions of items are common. }}
\end{abstract}

\maketitle

\section{Introduction}
\begin{figure}[tb]
    \resizebox{0.9\linewidth}{!}{
    \includegraphics{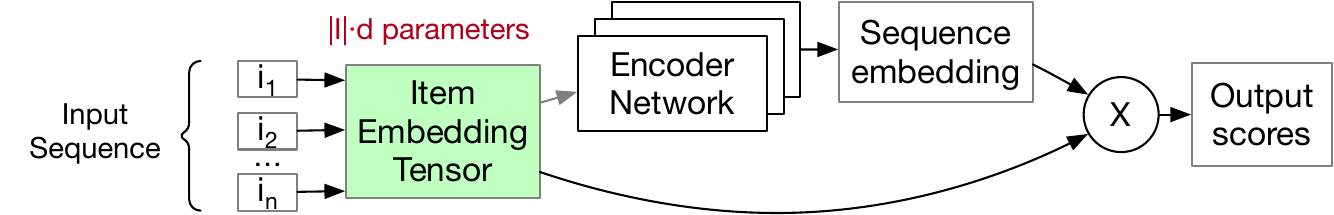}
    }
    \caption{Item embeddings in a typical sequential recommender system. These item embeddings are used in two ways: (i) to obtain sequence representation and (ii) to generate item scores. The embedding \sd{tensor} requires $|I| \cdot d$ trainable parameters, where $|I|$ is the items catalogue size, and $d$ is the size of an embedding. When catalogue size $|I|$ is large, item embeddings comprise most of the model's parameters.}\label{fig:embeddings_in_ach}
    \vspace{-1\baselineskip}
\end{figure}

\pageenlarge{3}
Sequential recommender systems are a class of recommendation models that use the sequence of user-item interactions to predict the next item. Most of the state-of-the-art models for sequential recommendation are based on deep neural networks, for example, recurrent neural networks~\cite{GRU4Rec, hidasiRecurrentNeuralNetworks2018}, convolutional neural networks~\cite{Caser,yuanSimpleConvolutionalGenerative2019}, and most recently, transformers~\cite{SASRec, BERT4Rec, PetrovRSS22, Bert4RecRepro}. All these models use learnable \emph{item embeddings} as an essential component in their model architectures. Figure~\ref{fig:embeddings_in_ach} illustrates item embeddings in a typical neural sequential recommendation model. As the figure shows, item embeddings usually have two roles in the model architecture: (i) to convert the sequence of input item ids to a sequence of item representation vectors and (ii) to convert the sequence embedding produced by the model into the distribution of predicted item scores. In both cases, a recommender system that works with an item set $I$ requires an embedding \sd{tensor} with $|I| \cdot d$ parameters, where $d$ is the size of each embedding. 

\pageenlarge{3}
\looseness -1 \craig{When a recommender has many items in the catalogue, various challenges arise in training the neural recommendation model. Firstly, the item embedding tensor \sff{may contain more model parameters than the rest of the model}. For example, there are more than 800 million videos on \crcs{YouTube~\cite{youtube_videos}. If a recommender model uses} 128-dimensional embeddings, the whole item embeddings tensor will have more than 100 billion parameters, which is comparable with the number of parameters of the largest available machine learning models~\cite{brownLanguageModelsAre2020}, even without accounting for the parameters of the model's intermediate layers. \crcs{\swsdm{This is a problem specific to recommender systems: in the related area of dense passage retrieval~\cite{karpukhinDensePassageRetrieval2020,khattabColBERTEfficientEffective2020}, passage embeddings are obtained by encoding passage text using a pre-trained language model; however, item side information,} such as text, is not necessarily available in a typical recommender systems scenario; therefore, item embeddings should be directly learned from the interactions}. Secondly, a large \sff{number} of such trainable parameters also makes the model prone to overfitting. 
\sd{A third challenge caused by the large catalogue is the size of the output scores tensor (rightmost tensor in Figure~\ref{fig:embeddings_in_ach}): for example, in BERT4Rec, it contains a score for each item for every position, for every sequence in the training batch, so training BERT4Rec with more than 1 million items in the catalogue may be prohibitively expensive~\cite{PetrovRSS22}. This problem is typically solved using negative sampling, whereby instead of computing the full output tensor, the model computes scores for a small proportion of negative items \crc{not interacted by the user} (e.g. SASRec~\cite{SASRec} uses one negative per positive). However, negative sampling comes with its own challenges (for example, it usually requires informative negative mining~\cite{rendle_item_2022}). Nevertheless, negative sampling is an orthogonal research direction, and in this paper, we use SASRec in cases where negative sampling is necessary.} \crc{We refer \crcb{the reader} to our recent publication~\cite{petrovGSASRecReducingOverconfidence2023}, where we analysed \crcb{the} negative sampling problem in details}. To summarise the challenges, a large embedding \sff{tensor} increases the model size, slows model \sd{training} down, and can necessitate further modelling tricks such as negative sampling, which bring their own challenges.}
 
 There are some existing methods\sdt{~\cite{xiaEfficientOnDeviceSessionBased2023, wangCompressingEmbeddingTable2022, kangLearningMultigranularQuantized2020a}} for item embedding compression (we discuss these methods in Section~\ref{sec:related}). However, most of these methods compress the embedding tensor \emph{after} the model is fully trained (including training the full embedding \sff{tensor}). However, as argued above, training may be prohibitively expensive in large-\crcb{catalogue} recommender systems.  Hence, this paper addresses the problem of a large item embedding tensor in sequential recommendation models \emph{at the training stage}.

\looseness -1 To mitigate this problem, we propose a novel RecJPQ technique inspired by the success of a recent Joint Product Quantisation (JPQ) work~\cite{zhanJointlyOptimizingQuery2021} for \crcs{passage} retrieval. JPQ itself is based on Product Quantisation (PQ)~\cite{jegouProductQuantizationNearest2011}, a popular method of compressing vectors by splitting them into sub-embeddings and encoding them using a discrete \emph{\crcb{item}\footnote{\crcs{For consistency, we explain prior work using \emph{item ids} instead of \emph{passage ids}.}} codebook} (the codebook maps from item ids to the associated \emph{\crcb{sub-item ids}}; see Section~\ref{ssec:pq} for the details). The main innovation of JPQ compared to the standard PQ method is that it learns the \emph{\crcb{sub-item} \sff{embeddings}} as part of the overall model training process. In contrast, PQ requires training the model first and only then compressing the embeddings \sd{(frequently, this second step uses external tools, such as FAISS~\cite{FAISS})}. This means that JPQ does not need to keep the embedding matrix in memory during model training. We argue that this innovation is valuable for recommender systems. Indeed, as mentioned above, real-life recommender systems can have hundreds of millions of items in their catalogues and keeping full embeddings \sn{tensor} in memory may be prohibitively expensive. This is particularly important for deep-learning-based sequential recommender systems because these models require keeping the whole model in GPU (or TPU) memory during training. GPU memory is costly even when compared to regular computer RAM. 

\pageenlarge{3}
\looseness -1 Unfortunately, it is hard to adapt  JPQ to the recommendation scenario, as it is specific to textual information retrieval. In particular, JPQ assumes the existence of a pre-trained \swsdm{\crcs{(passage retrieval)} model and index}, which it uses to assign items to \crcb{sub-item ids} (see more details in Section~\ref{ssec:jpq}). These pre-trained models rarely exist in item recommendations. Hence, in RecJPQ, we \craig{experiment with performing the initial assignment of \crcb{sub-ids}}  using three different strategies: \sd{(i) discrete truncated  SVD (\crcb{sub-ids} obtained by discretising \craig{the item representations obtained by an SVD decomposition of the user-item matrix}), (ii) discrete BPR (\crcb{sub-ids} obtained by discretising the item embeddings obtained from BPR) and (iii) random assignments}. We describe \craig{these assignment strategies} in detail in Section~\ref{sec:recjpq}.

RecJPQ is a model component that replaces traditional item embeddings in sequential recommender systems. \sd{In general, it can be applied to any recommender system based on item embeddings, but in 
 this paper, we focus specifically on sequential models, as in these models, item embeddings comprise the biggest part of the model (e.g. sequential models usually do not have user embeddings)}. In contrast with existing methods, RecJPQ does not require training full uncompressed embedding and does not modify the original model loss function. Our experimentation on three datasets (see Section~\ref{sec:experiments}) demonstrates that RecJPQ can be successfully applied to different models, \sn{including SASRec~\cite{SASRec}, BERT4Rec~\cite{BERT4Rec} and GRU~\cite{GRU4Rec, PetrovRSS22}}, achieving a large factor of embeddings compression (e.g.\ \sd{47.94$\times$ compression of SASRec on Gowalla}) \craig{without any effectiveness degradation}. Moreover, \sd{on 2 out of \craig{our} 3 experimental datasets,} applying RecJPQ \emph{increases} model performance (e.g. \sd{+0.96\% NDCG@10 on Booking.com dataset, significant improvement}); we attribute these improvements to model regularisation. 

\sn{In short, the contributions of this paper are as follows:}
\crcf{
        (1) we propose RecJPQ, a novel technique for reducing the size of sequential recommendation models during training based on Joint Product Quantisation;
        (2) we propose three strategies for assigning \crcb{sub-item ids} to items, two of which (discrete truncated SVD and discrete BPR) assign similar codes to similar items, and one assigns codes randomly; 
        (2) we perform an extensive experimental evaluation of RecJPQ on three datasets and show that RecJPQ allows reducing the models' size without hindering the model performance. 
}
\sn{The rest of the paper is organised as follows: Section~\ref{sec:related} introduces related work on embeddings compression and identifies limitations of existing methods; Section~\ref{sec:pq_jpq} covers Product Quantisation (PQ) and Joint Product Quantisation (JPQ) -- the methods, which serve as the basis for our work; Section~\ref{sec:recjpq} introduces RecJPQ and covers \crcb{sub-id} assignment strategies for RecJPQ; in Section~\ref{sec:experiments} we experimentally evaluate RecJPQ; Section~\ref{sec:conclusion} contains final remarks.}

\pageenlarge{3}
\vspace{-0.5\baselineskip}
\section{Related work}\label{sec:related}
\begin{table}[tb]
\caption{Existing embedding compression methods. \crcs{The} desired method characteristics are highlighted in bold.}\label{tb:model_compression}
\vspace{-0.7\baselineskip}
\resizebox{\linewidth}{!}{
  \begin{tabular}{lllllll}
\toprule
\makecell[l]{Model\\ Agnostic} & Method & Backbone & \makecell[l]{Sequential\\ backbone} & \makecell[l]{Trains full\\ embeddings} & \makecell[l]{Compression \\ unit} \\
\midrule
No & EODRec~\cite{xiaEfficientOnDeviceSessionBased2023} & SASRec~\cite{SASRec} & \textbf{Yes} & Yes & \textbf{Item} \\
 & LightRec~\cite{lianLightRecMemorySearchEfficient2020} & DSSM~\cite{huangLearningDeepStructured2013} & No & Yes & \textbf{Item} \\
 & MDQE~\cite{wangCompressingEmbeddingTable2022} & SASRec~\cite{SASRec} & \textbf{Yes} & Yes & \textbf{Item} \\ \midrule
\textbf{Yes} & PreHash~\cite{shiUserEmbeddingMatrix2020} & BiasedMF~\cite{KorenMF}; NeuMF~\cite{heNeuralCollaborativeFiltering2017} & No & \textbf{No} & User \\
 & Quotient Remainder~\cite{shiCompositionalEmbeddingsUsing2020} & DCN~\cite{wangDeepCrossNetwork2017}; DLRM~\cite{naumovDeepLearningRecommendation} & No & \textbf{No} & Features \\
 & MGQE~\cite{kangLearningMultigranularQuantized2020a} & SASRec~\cite{SASRec}; NeuMF~\cite{heNeuralCollaborativeFiltering2017}; GCF~\cite{heNeuralCollaborativeFiltering2017}; & \textbf{Yes} & Yes & \textbf{Item} \\ \midrule
\textbf{Yes} & RecJPQ (ours) & SASRec~\cite{SASRec}; BERT4Rec~\cite{BERT4Rec}; GRU~\cite{GRU4Rec} &  \textbf{Yes} & \textbf{No} & \textbf{Item} \\
\bottomrule
\end{tabular}

}
\vspace{-1\baselineskip}
\end{table}
\looseness -1 \sd{This section covers existing work on compressing and discretising embeddings in recommender systems, \sd{identifies the limitations in existing work and positions our contributions in the context of existing methods}. Table~\ref{tb:model_compression} summarises existing methods and positions RecJPQ, our proposed compression technique.} The table highlights with boldface the desirable characteristics necessary for training a large-scale\footnote{\sd{For simplicity, we say that a catalogue is "large-scale" if it has more than 1 million items, as it becomes challenging to train recommender systems on that scale~\cite{PetrovRSS22}.}} sequential model, specifically: the method can be applied to work with different \craig{{\em backbone} sequential models}, and \craig{does not require training} full embeddings (as we work with the assumption that \crcb{the} full embeddings tensor does not fit into GPU memory); and we want the model to focus on item embeddings rather than embeddings of other entities, such as users or features. As \craig{illustrated in the table}, the methods for compressing the models can be broadly divided into two groups: \emph{model \craig{dependent}} and \emph{model agnostic}.

\looseness -1 In the model-dependent methods~\cite{lianLightRecMemorySearchEfficient2020, xiaEfficientOnDeviceSessionBased2023}, \crcb{the} embedding compression \sn{mechanism \craig{is} integrated as a component into the recommendation model itself}. \craig{Hence\crcb{,} the training architecture of these methods} has to be aware of the compression, and the loss function includes components responsible for the embedding compression. For example, LightRec~\cite{lianLightRecMemorySearchEfficient2020} uses the \sn{Deep Semantic Similarity Model (DSSM)}~\cite{huangLearningDeepStructured2013} as the backbone model and uses an additional knowledge distillation component in the loss, which allows for learning compressed representations of the embeddings. \sdt{However, DSSM is not a sequential model, and it is unclear whether or not the method can be adapted to the sequential recommendation \crcb{task}.} Similarly, \crcs{EODRec~\cite{xiaEfficientOnDeviceSessionBased2023}\crcb{,} uses SASRec~\cite{SASRec}} as its backbone, one of the most popular sequential models based on \craig{the} Transformer \craig{architecture}~\cite{Transformer}. The loss function of EODRec also consists of four components, some of which are responsible for recommendation, and others are responsible for embedding compression. \sdt{\crcs{While EODRec's backbone (SASRec) is an efficient model,} in many cases, other models such as BERT4Rec \crcb{exhibit} better results~\cite{BERT4Rec, Bert4RecRepro}. In general,}  while some model-dependent methods \crcs{may reduce the size} of a trained model, \sn{these methods have several limitations, which make them unsuitable for training sequential recommendation models with large catalogues. In particular:}

    \labeltext{\textbf{L1}}{limitation:model_tied} Model-dependent \sn{methods} are, \craig{by their nature}, tied to a specific model, making them inflexible when adapting to a specific task. For example, LightRec~\cite{lianLightRecMemorySearchEfficient2020} uses a non-sequential DSSM model as a backbone. \sn{The core component of LightRec (Recurrent Composite Encoding) is tightly integrated into \craig{the} DSSM architecture, and it is unclear whether or not it can be used outside of DSSM.}

    \pageenlarge{3}
    \labeltext{\textbf{L2}}{limitation:full_embedding} Model-dependent methods usually require training (uncompressed) item embeddings and then use knowledge distillation or teacher-student techniques to obtain compressed representations of the embeddings. This approach substantially reduces the final model size, \craig{thereby} \sn{helping inference on smaller devices}, but requires a large amount of GPU memory while training, \sn{thereby} limiting the overall number of items in the catalogue. For this reason, the main positioning for EODRec model~\cite{xiaEfficientOnDeviceSessionBased2023} is the on-device recommendation: while the final model produced by this method is small, it requires storing full item embeddings while training. \swsdm{Post-training quantisation methods~\cite{yaoZeroQuantEfficientAffordable2022a}, which recently became popular to reduce the size of large language models via quantising their weights into lower-precision numbers (e.g.\ float16, or int8) also have this limitation -- they need to have access to \crcb{the} full model before quantising. Similarly, Mixed Precision Training~\cite{micikeviciusMixedPrecisionTraining2018} builds a smaller precision model, but it requires keeping full precision weights in memory. Placing the embeddings tensor into Approximate Nearest Neighbours~\cite{FAISS} or Hierarchical Navigable Small Worlds~\cite{malkovEfficientRobustApproximate2020} indexes also requires access to the full embedding tensor at model training time, and therefore also exhibits this limitation.} 
    
    \looseness -1 \labeltext{\textbf{L3}}{limitation:complex_loss} Model-dependent methods require multi-component loss functions, some of which are responsible for the recommendation task and others for the model compression. 
    \sn{This is a form of multi-objective optimisation, which is a challenging problem~\cite{sener_multi-task_2018, deb_multi-objective_2014}}\sdt{, as finding the balance between the loss components for different objectives usually requires extensive hyperparameters search}. 
    
On the other hand, \craig{the existing} model-agnostic methods~\cite{shiUserEmbeddingMatrix2020, shiCompositionalEmbeddingsUsing2020} do not depend on the specific model architecture, and \craig{likewise} do not add extra components to the loss functions. Typically, these methods implement a mechanism that takes the place of the embeddings \sdt{tensor} in the backbone model, and \craig{hence} can be used with many models. 
However, on inspection of the relevant work, we \crcb{elicit} additional limitations of these methods: 

   \labeltext{\textbf{L4}}{limitation:not-similar} Many methods are not designed for compressing item embeddings. For example, PreHash~\cite{shiUserEmbeddingMatrix2020} is a method specific for compressing user embeddings (i.e.\ it uses the user's history to construct user embeddings). \sn{The method uses an attention network over the history of user interactions. Adapting this network \sdt{structure} for items \sdt{embeddings} is a hard task: a user may only interact with a few items; in contrast, a popular item may \craig{be interact with by} millions of users. The attention mechanism depends quadratically on the sequence length, and therefore, applying it to users who interacted with a popular item \craig{would} be prohibitively expensive.}

  \labeltext{\textbf{L5}}{limitation:not-items} Finally, many methods lack structure in compressed item representations. This leads to situations \craig{where} unrelated items have similar representations and, \craig{conversely,} when similar items have \craig{dissimilar} representations. Both these cases may limit the models' generalisation ability and hinder the models' performance. \sd{One example of \craig{such unstructured} methods is \swsdm{hashing-based} Quotient Remainder~\cite{shiCompositionalEmbeddingsUsing2020}, which \swsdm{compresses} embeddings of categorical features (e.g., genres). Quotient Remainder assigns feature codes based on the quotient and the remainder of the division of the feature id by some number.} \sd{When applied to item ids (items can also be seen as categorical features), the quotient and the remainder are unrelated to the item characteristics. Hence, similar items are unlikely to have similar codes}. \sn{Nevertheless, Quotient Remainder is one of the few methods that can be used to train a \sd{model on a large-scale dataset}, and therefore, we use this method as a baseline in our experiments}.

\pageenlarge{3}
\looseness -1 Overall, \sdt{among the} related work, we \sn{argue} that existing methods \craig{exhibit} \craig{a number} of Limitations (L1-L5). \craig{In summary, the} model-dependent methods require training full embeddings first, limiting the maximum number \craig{of items that can be considered in the catalogue of the recommender system}. On the other hand, methods which do not require training full embeddings first, such as Quotient Remainder, rely on heuristics and may result in unrelated items having similar representations. \swsdm{On the surface, the nearest related work to ours is VQ-Rec~\cite{houLearningVectorQuantizedItem2023a} as it also applies JPQ-style technique to sequential recommendation; however, similarly to JPQ, it relies on the availability of textual features and pre-trained language models. In contrast, our work's main novelty is adapting JPQ to the scenario when (e.g., textual) side information is \crcb{not} available.} \sn{In the next section, we describe \crcb{Product Quantisation -- a vector compression method and } JPQ~\cite{zhanJointlyOptimizingQuery2021}, a method of embedding compression for information retrieval, which directly learns embeddings in compressed form, reducing GPU memory requirements. Then, in Section~\ref{sec:recjpq}, we propose RecJPQ -- an adaptation of JPQ to the sequential recommendation task, which successfully addresses Limitations L1-L5.}

\begin{table}[tb]
  \caption{Analysis of PQ's impact on memory requirements for storing item embeddings tensor for selected recommendation datasets, based on 512-dimension float32 vector embeddings. The table compares base memory usage with different code lengths, shown as percentages relative to the base.}\label{tb:centroids_memory}
  \vspace{-0.7\baselineskip}
  \resizebox{\linewidth}{!}{\begin{tabular}{l|r|r|rrr}
\toprule
\multirow{4}{*}{Dataset} & \multirow{4}{*}{Num Items} & \multicolumn{4}{c}{\swsdm{Size of Item Embedding Tensor}} \\
 \cline{3-6}& & \makecell[r]{Base} & 
\makecell[r]{Code length=2 \\ 512 centroids\\ 1.00 MB} & 
\makecell[r]{Code length=8 \\ 2,048 centroids\\ 4.00 MB} & 
\makecell[r]{Code length=32 \\ 8,192 centroids\\ 16.00 MB} \\
\midrule
MovieLens-1M & 3,416 & 6.67 MB & 14.988\% & 59.953\% & 239.813\% \\
Booking.com & 34,742 & 67.86 MB & 1.474\% & 5.895\% & 23.580\% \\
Gowalla & 1,280,969 & 2.44 GB & 0.040\% & 0.160\% & 0.640\% \\
\bottomrule
\end{tabular}}
  \vspace{-1.3\baselineskip}
\end{table} 

\vspace{-0.5\baselineskip}
\section{\sn{Product Quantisation and JPQ}} \label{sec:pq_jpq}

\crcs{We now describe Product Quantisation (PQ) and Joint Product Quantisation (JPQ), two methods \crcb{that} serve as \crcb{backbones} for our method. Section~\ref{ssec:pq} covers PQ, a classic embedding compression technique. Section~\ref{ssec:jpq} describes JPQ -- a recently proposed information retrieval method that learns compressed embeddings directly instead of compressing them after the model training.}

\subsection{Product Quantisation}\label{ssec:pq}
\looseness -1 \emph{Product quantisation}~\cite{grayVectorQuantization1984, jegouProductQuantizationNearest2011} is a \sd{well-cited} method of compressing vectors \sd{used by many libraries, such as FAISS~\cite{FAISS} \& nanopq~\cite{nanopq}}. \swsdm{Its} main idea is to split a collection of $d$-dimensional vectors, $V$, into $m$ collections $V_{i}; i=1..m$ of smaller vectors of $\frac{d}{m}$ dimensions each. The original vectors can be recovered back via concatenation: $V = concat(V_1, V_2, ... V_m)$. Product quantisation then clusters each $V_i$ into $b$ clusters (e.g.\ using the k-means algorithm~\cite{macqueenMethodsClassificationAnalysis1967}) and replaces each vector $v_{ij}$ with the centroid of the assigned cluster $c_{ij} \approx v_{ij}$. \swsdm{With this replacement}, the original vectors collection can be approximated as a concatenation of the \crcf{\emph{sub-vector matrices}} \sn{$C_1 ... C_m$, which are constructed from $V_1 ... V_m$ by replacing each vector $v_{ij}$ by the closest \crcb{cluster centroid} $c_{ij}$}: $V \approx concat(C_1, C_2, ... C_m)$. In each \crcb{sub-vector} matrix $C_i$, there are, at most, $b$ different rows, as each row corresponds to one of the centroids of the clusters, so instead of storing full matrix $C_i$, we can store these unique rows in the separate tensor $\hat{C}_i$, \craig{whose} elements $\hat{c}_{ij}, j\in\{1 .. b\}$ \sn{correspond to the unique \crcb{sub-vectors}. \crcf{To compress the approximate embedding matrix}}, we need only store the ids of \crcs{the} \crcb{sub-vectors (sub-ids) \crcs{for each item}}. Overall, there are $m$ \crcb{sub-vector} sets $C_i$, so each vector can be encoded using $m$ integer codes, and \sn{each code} can have $b$ different values; therefore, overall, this scheme can encode up to $b^m$ different vectors. 
\sn{The vector of \crcb{sub-item} ids $g_i = \{g_{i1}, g_{i2} ..., g_{im}\}$ associated with the vector $v_i \in V$ is known as the \emph{code} of the vector $v_i$~\cite{jegouProductQuantizationNearest2011}. The number \crcb{$m$ of sub-item ids} associated with \crcb{each} item \crcb{is called} the \emph{length} of the code. The table $G$ of codes associated with each vector from $V$ is also known as a \emph{codebook}~\cite{jegouProductQuantizationNearest2011}}. 

\begin{figure}[tb]
    \vspace{-1\baselineskip}
    \includegraphics[width=0.5\textwidth]{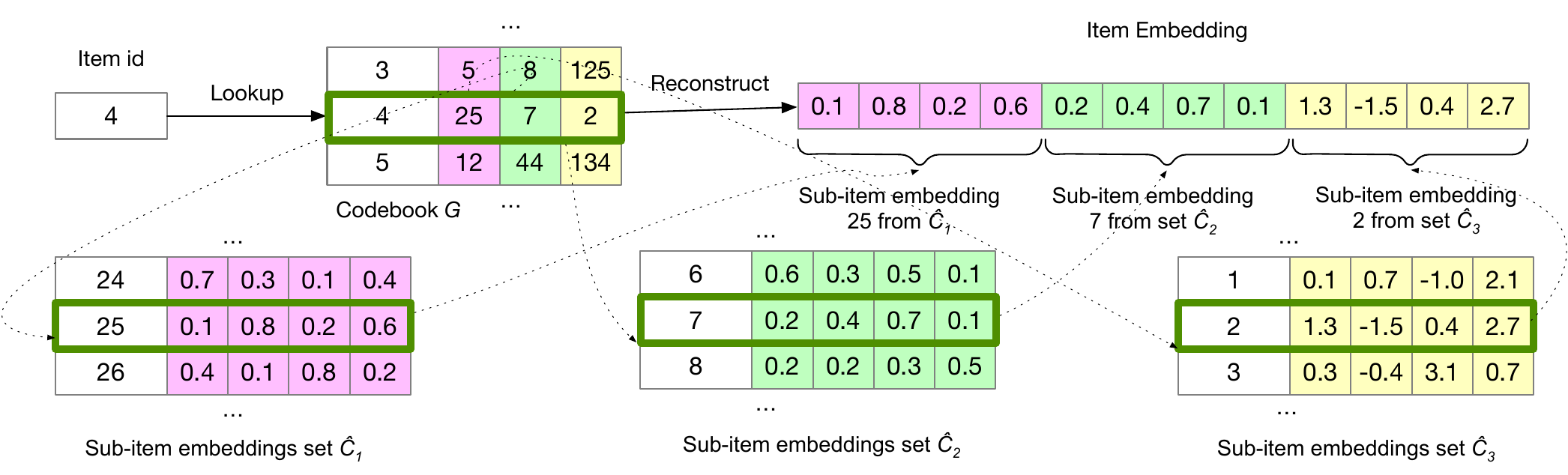}
    \caption{Reconstruction of item embeddings \crcs{using Product Quantisation}: Codebook length $m=3$, item embedding length $d=12$, number of \crcb{sub-ids} per split $b=256$.}
    \label{fig:embedding_reconstruction}
    \vspace{-2 \baselineskip}
\end{figure}

\looseness -1 Figure~\ref{fig:embedding_reconstruction} illustrates the vector reconstruction process \craig{applied} to item embeddings. For each \sn{\crcb{sub-item} id $g_{ij}$} in the codes vector $g_i$ of an item $i$, we extract \crcb{a sub-item embedding} associated with this \sn{sub-item id} and then concatenate the \crcb{sub-item embeddings} to obtain \sdt{reconstructed} item embeddings.

\pageenlarge{3}
\looseness -1 The number of splits $m$ is usually considerably smaller than the original vector dimensions $d$: $m \ll d$ to achieve compression.
The number of \crcb{sub-embeddings} per split, $b$, is typically a power $k$ of 256, so \craig{that} the codes can be stored as $k$-byte integers. In this paper, for simplicity and following JPQ~\cite{zhanJointlyOptimizingQuery2021}, we fix $k=1$, so each \crcb{sub-item id} can be represented \craig{with a single} byte; therefore, we only store $m$ bytes for each item.  \swsdm{Even with fixed $k$ (and therefore $b$), we can adjust model capacity by controlling the number of \crcb{sub-item ids} associated \crcb{with} each item, $m$, and the dimension of the \crcb{sub-item embeddings}, $\frac{d}{m}$, by adjusting $d$.} \swsdm{Figure~\ref{fig:embedding_reconstruction} illustrates \crcs{PQ applied to item embeddings} for $m=3$, $d=12$, $b=256$. Further, to illustrate the achieved compression}, if embeddings are stored as 256-dimensional float32 vectors, a full-item embedding requires 1KB of memory. After compression using product quantisation with $m=8$ splits, we only need to store 8 bytes per item (0.78\% of the original memory requirement). While some memory is also required to store the \crcb{sub-item embeddings} themselves, this is negligible for large datasets compared to the original vector requirements (see also Table~\ref{tb:centroids_memory} for an analysis of \crcb{sub-item embeddings} memory requirements).

\looseness -1 Product Quantisation is a well-established vectors compression technique, which \sn{has been shown} to be successful in approximate nearest neighbour methods~\cite{jegouProductQuantizationNearest2011, FAISS, geOptimizedProductQuantization2014, liApproximateNearestNeighbor2020}, information retrieval~\cite{khattabColBERTEfficientEffective2020, santhanamColBERTv2EffectiveEfficient2022, izacardMemoryEfficientBaseline2020} and recommender systems~\cite{balenPQVAEEfficientRecommendation2019, kangLearningMultigranularQuantized2020a, wangCompressingEmbeddingTable2022}.
However, Product Quantisation requires the \emph{full} embeddings \sn{tensor} to be trained before the embeddings are compressed. Indeed, the quantisation operation is not differentiable, and therefore, the model can not be trained end-to-end. Therefore, it requires first training full (non-quantised) model and only \crcb{then applying compression}. Some recommendation models (e.g.~\cite{kangLearningMultigranularQuantized2020a}) use differentiable variations of Product Quantisation to allow end-to-end training, but these methods still require training full embeddings alongside the quantised versions. 

\sn{Overall, Product Quantisation addresses Limitations~\ref{limitation:model_tied} (it is model agnostic),~\ref{limitation:not-items} (it is applicable for training item embeddings compression), and~\ref{limitation:not-similar} (when \crcb{sub-item ids are} assigned using clusterisation, similar items will have similar codes). However, it does not address Limitations~\ref{limitation:full_embedding} (it requires training full embeddings first)\crcb{, and~\ref{limitation:complex_loss}} (PQ uses a separate loss function for embeddings reconstruction, which is not aligned \crcb{with ranking loss}). In the next section, we discuss Joint Product Quantisation, a method \swsdm{that can be adapted to} addressing these remaining limitations.}

\vspace{-0.5\baselineskip}
\subsection{Joint Product Quantisation} \label{ssec:jpq}
\looseness -1 Zhan et al.\ recently proposed \emph{Joint Product Quantisation (JPQ)}~\cite{zhanJointlyOptimizingQuery2021}, a Product Quantisation-based method developed for dense information retrieval. The main difference with the classic product quantisation is that JPQ generates item codes \emph{before} training the model. As code assignment is the only non-differentiable operation in Product Quantisation, therefore, when the codes are assigned before training the model, the model can be trained end-to-end without training full item embeddings -- JPQ essentially replaces the embeddings \sd{tensor} in the model, where item embeddings are constructed via \crcb{sub-item \crcs{embeddings}} concatenation, as illustrated in Figure~\ref{fig:embedding_reconstruction}. Assuming that the codebook in the figure is a constant, all other parameters can be learned using standard gradient descent \crcf{on model's loss. In particular, this} means that JPQ does not require special loss function components to learn \crcb{sub-item embeddings}.
\pageenlarge{3}
\looseness -1 \sn{Compared to the original Product Quantisation method, JPQ addresses Limitations~\ref{limitation:full_embedding} (it does not require training full item embeddings) and~\ref{limitation:complex_loss} (it does not require a specific loss function). However, in contrast to \craig{plain} PQ, JPQ does not provide a mechanism to assign similar embeddings to similar items and, therefore, does not address Limitation~\ref{limitation:not-similar}. The \crcb{sub-item id} assignments method} proposed in the original JPQ paper is specific for text retrieval \sd{(as it relies on the existence of a pre-built index for a text document collection generated using the STAR model~\cite{zhan_optimizing_2021})}. \sd{In the next section, we introduce RecJPQ, an adaptation of JPQ to the sequential recommendation scenario, which does not rely on text-specific datasets and models.} \swsdm{Our adaptation of JPQ to sequential recommendation requires careful design of novel \crcb{sub-item id} assignment strategies. Indeed, to the best of our knowledge, this is the first adaptation of the JPQ method to sequential recommendation.}

\vspace{-0.5\baselineskip}
\section{R\MakeLowercase{ec}JPQ} \label{sec:recjpq}
\emph{RecJPQ} is a Joint Product Quantisation-based method for training recommendation models with a large catalogue of items. As we discuss in Section~\ref{ssec:jpq}, the method used by JPQ for initial \crcb{sub-item id} assignments relies on the existence of a pre-built index of documents and, therefore, can not be directly used for recommendation scenarios. Hence, the main difference between RecJPQ and the original JPQ is \crcb{sub-item id} assignment strategies. %
\sd{In general, RecJPQ can be described as performing the following steps:}

\crcf{\textbf{Step 1. }}Build the item-code mapping matrix (codebook) using one of the \crcb{sub-item id} assignment strategies (\crcf{see} Section~\ref{ssec:centroid_assignment_strategies}). 
 
\crcf{\textbf{Step 2. }} Initialise the \crcb{sub-item} embeddings randomly.

\crcf{\textbf{Step 3. }} Replace the item embedding tensor with the concatenation of \crcb{sub-item embeddings} associated with each item).

\crcf{\textbf{Step 4. }} Train the model end-to-end using the model's original training task and loss function (this process also trains the \crcb{sub-item} embeddings, so they do not need to be trained separately).  

\sd{The way RecJPQ builds the codebook before training the main model is also similar to how language models (such as BERT~\cite{BERT}) train a tokenisation algorithm before training the main model. Language models also learn embeddings of sub-word tokens instead of learning embeddings of full words to reduce the model's size; similarly, RecJPQ learns \crcb{sub-item embeddings} instead of learning embeddings of full items.} \swsdm{Below, we describe \crcb{sub-item id} assignment strategies used by RecJPQ}.

\pageenlarge{3}
\vspace{-1\baselineskip}
\subsection{\crcb{Sub-Item Id} Assignment Strategies}\label{ssec:centroid_assignment_strategies}
 
\crcf{\textbf{{Random \crcb{Sub-Item Id} Assignments.}}}
In the most simple scenario, we can assign items to \crcb{sub-item ids} randomly. We compose the item code out of $m$ random integers in this case. RecJPQ with random \crcb{sub-item id} assignments strategy does not address Limitation~\ref{limitation:not-similar} (similar items do not have similar codes). Indeed, with random \crcb{sub-item id} assignments, RecJPQ becomes similar to other "random" embeddings compression methods, such as the hashing trick~\cite{weinberger_feature_2010} or \crcs{the} Quotient Reminder method~\cite{shiCompositionalEmbeddingsUsing2020}. 
The main problem with these methods is that unrelated methods are forced to share parts of their representation, which limits the \crcb{generalisability} of the models. However, as we show in Section~\ref{sec:experiments}, sometimes random assignments may be beneficial, as the random assignments strategy acts as a form of regularisation. 
\looseness -1 Nevertheless, as the random \crcb{sub-item id} assignments strategy does not address Limitation~\ref{limitation:not-similar} \sdt{(similar items should have similar representations)}, we introduce further \crcb{sub-item id} assignment strategies that can address this limitation in the next sections. 

\crcf{\textbf{Discrete Truncated SVD.}} %
\looseness -1 As discussed in Section~\ref{ssec:jpq}, \sn{the only limitation not addressed by JPQ is Limitation~\ref{limitation:not-similar} (similar items should have similar codes). Random \crcb{sub-item id} assignments, as discussed \crcf{above}, do not address this limitation either. Hence, in this section, we design a \crcb{sub-item id} assignments method that addresses this remaining limitation and assigns similar codes to similar items. Some approaches have used side information, such as textual data for item representations~\cite{rajput_shashank_recommender_nodate}; however, we address the more generic classic sequential recommendation scenario \crcf{and, therefore,}} must infer item similarities from the user's sequences. To \crcf{do} this, we employ the  SVD algorithm, which has been shown to achieve good results in learning item representations~\cite{zhou_personalized_2012} for recommender systems.

\looseness -1 We first compute \crcb{a} matrix of sequence-item interactions $M$, where rows correspond to sequences (users), and columns correspond to items. This matrix's elements $m_{ij}$ are either 1 if $i^{th}$ contains interactions with item $j$ and 0 otherwise. We then compute the truncated SVD decomposition of matrix $M$ with $m$ latent components:
    $M \approx U \times \Sigma \times V^T$, 
where $U$ is the matrix of user embeddings, $V$ is the matrix of item embeddings, and $\Sigma$ is the diagonal matrix of largest singular values. 

\looseness -1 Our initial experiments showed that some items have equal embeddings after performing truncated SVD decomposition. \crcs{This occurs when two items have been interacted with by exactly the same set of users.} To ensure that all items have different embeddings, we normalise $V$ using the min-max normalisation  range and add a small amount of Gaussian noise: 
    $\hat{v}_{ab} = \frac{v_{ab} - \min_{k}{v_{ak}}} {\max_{k}{v_{ak}} - \min_{k}{v_{ak}}} + \mathcal{N}(0, 10^{-5}); \forall v_{ab} \in{V}$
The variance of the noise ($10^{-5}$) is negligible compared to the range of possible values of normalised embeddings ($[0..1]$ after min-max normalisation). Therefore it has a very small influence on the position of the items in the embeddings space. However, if two items have exactly the same embeddings after decomposition (e.g.\ this can happen if two items appear in exactly the same set of sequences), the noise allows us to distinguish these two embeddings.  
Lastly, \craig{the assignment of \crcb{sub-item id} involves discretising} each dimension of the normalised item embeddings into $b$ quantiles so that each quantile contains an approximately equal number of items. 
We use these bins as \crcb{sub-item id} assignments for the items. Note that although this method requires computing \sdt{an $m$-dimens\-i\-onal} item embeddings \sdt{tensor} (and there can be hundreds of millions of items), it does not require computing them as part of a deep learning model training on a GPU. Indeed, truncated SVD is a well-studied problem. There are effective algorithms for performing it that do not require modern GPUs~\cite{halko_finding_2011}. Moreover, as the method only requires computing $m$-dimensionsional embeddings, \craig{the table will be} many times smaller than full $d$-dimensional embeddings, so the method requires $\frac{d}{m}$ times less memory to store embeddings. 
Finally, performing truncated SVD is possible in a distributed manner \crcs{(e.g., using Apache Spark~\cite{spark_svd})}, which \crcf{allows} truncated SVD \crcs{to be applied} on large datasets.
\pageenlarge{3}
In summary, discrete truncated SVD allows assigning similar codes to similar items; it does not require a GPU for intermediate computations and can be easily performed for very large datasets. 

\looseness -1 \crcf{\textbf{Discrete BPR.}}
Truncated SVD is not the only Matrix Factorisa\-t\-ion method that can be used for initial \crcb{sub-item id} assignments. In particular, we also use the classic BPR approach~\cite{rendle_bpr_2009} to obtain coarse item embeddings. The method also learns user embeddings (or, in our case, sequence embeddings) $U$ and item embeddings $V$. The \craig{estimate of the relevance} of an item $i$ for user $j$ is defined as the dot product of user and item embeddings: \sd{$r = u_j \cdot v_i$}. In contrast with truncated SVD, BPR does not directly approximate the user-item interaction matrix. 
Instead, BPR optimises a pairwise loss function that aims to ensure that positive items are scored higher than negative items: $\mathcal{L}_{BPR} = -\log(\sigma (u_i \cdot v_{j^+} - u_i \cdot v_{j^-}))$,
where $v_{j^+}$ is the embedding of a positive item for the user $u$, $v_{j^-}$ is the embedding of a randomly sampled negative item, and $\sigma$ is the sigmoid function.

\crcb{BPR is one of the most cited methods in recommender systems}; therefore, we use BPR as an alternative strategy for coarse item embedding learning. The rest of the discrete BPR strategy is the same as in the truncated SVD: we also normalise the learned embeddings using \crcb{a} min-max normalisation and add a small amount of Gaussian noise to ensure different embeddings for different items \crcb{are obtained}. Similar to truncated SVD, BPR does not require learning on a GPU, and there exist distributed implementations~\cite{spark_bpr} that \craig{allow for} learning item embeddings on very large datasets, so overall, it can be used as an alternative to truncated SVD for \crcb{sub-item id} assignments in RecJPQ.

This concludes the description of the \crcb{sub-item id} assignment strategies for RecJPQ. We now discuss why RecJPQ may act as \sd{a regularisation mechanism and} improve the performance on the datasets with many long-tail items. 

\vspace{-1\baselineskip}
\subsection{RecJPQ as a Regularisation Mechanism}\label{ssec:recjpq_is_reg}
\looseness -1 The interactions with items in recommender systems typically \crcb{exhibit a} long tail distribution~\cite{park_long_2008}, meaning that \crcb{a} few popular items \crcb{receive} the most interactions. In contrast, most items comprise the "long tail" with few interactions. 
As the training data for these long-tail items is limited, \crcb{recommendation} models \crcb{can} suffer from overfitting on \crcf{such items}~\cite {zhang_model_2021}, \crcf{causing} overall performance degradation. 

\looseness -1 Goodfellow et al.~\cite[\crcb{Ch.} 7]{goodfellow_deep_2016} argued that one of the most powerful \crcf{regularisation} techniques is~\emph{\crcb{parameter} sharing} -- a technique where certain parameters of the model are forced to be equal. RecJPQ is a special case of parameter sharing: we force different items to share parts of their embeddings. This prevents the model from \crcf{learning embeddings} \craig{that} are too specific to only a few training sequences, as each part of the embedding appears in many other sequences via the sharing mechanism.  In our experiments (Section~\ref{sec:experiments}), we indeed observe that RecJPQ may act as a model regulariser and improve the model's performance; this is especially \craig{apparent} in the Gowalla dataset, where the proportion of long-tail items is the largest.

\looseness -1 \crcf{\textbf{RecJPQ Summary}. To conclude Section~\ref{sec:recjpq}, we summarise} that  RecJPQ is a model component that takes the place of the item embeddings \sdt{tensor} in sequential recommender systems. RecJPQ is based on the JPQ method, which is a variation of the PQ method. RecJPQ addresses all of the limitations described in Section~\ref{sec:related}: it is model-agnostic (Limitation~\ref{limitation:model_tied});
does not require training full embeddings (Limitation~\ref{limitation:full_embedding}); does not modify the backbone model's loss function (Limitation~\ref{limitation:complex_loss}); it is suitable for item embeddings compression (Limitation~\ref{limitation:not-items}); it can assign similar codes to similar items with the help of discrete truncated SVD or discrete BPR (Limitation~\ref{limitation:not-similar}). \craig{Futhermore, we argue that RecJPQ may act as \sd{a} model regulariser, which is} an additional advantage when there are many long-tail items in the catalogue. 
This concludes the description of RecJPQ. In the next section, we experimentally evaluate RecJPQ and analyse its effects on required memory and on the model performance. 

\pageenlarge{3}
\vspace{-0.5\baselineskip}
\section{Experiments}\label{sec:experiments}
\begin{table}[tb]
    \centering
    \caption{Characteristics of the datasets. "Long tail~\%" is the percentage of items with less than 5 interactions.} \label{tb:datasets}
    \vspace{-0.5\baselineskip}
    \resizebox{0.8\linewidth}{!}{
    \begin{tabular}{lrrrrr}
    \toprule
    Dataset &  Users &  Iems &  Interactions &  \makecell[r]{\crcb{Avg. length}} & \makecell[r]{Long  tail \%}\\
    \midrule
    MovieLens-1M &       6,040 &       3,416 &            999,611 &           165.49 & 0.0\% \\
    Booking.com &   140,746 &      34,742 &           917,729 &            6.52  & 61.8\%\\
    Gowalla  &      86,168 &    1,271,638 &           6,397,903 &            74.24 & 75.8\%\\
    \bottomrule
    \end{tabular}}
    \vspace{-1\baselineskip}
\end{table}

Our experiments address the following research questions: 
\begin{enumerate}[font={\bfseries}, label={{\tiny \bfseries RQ\arabic*}}, leftmargin=*]
    \item What \crcs{is} the effect of \crcs{the} \crcb{sub-item id} assignment strategy?
    \item How do code length $m$ and embedding size impact effectiveness? 
    \item What is the effect of RecJPQ on size/effectiveness tradeoff?

\end{enumerate}
\vspace{-0.7\baselineskip}
\subsection{Experimental Setup}

\textbf{\crcf{Backbone Models.}}
\looseness -1 In our experiments, we use two state-of-the-art Transformer-based recommendation models: BERT4Rec~\cite{BERT4Rec} -- a model \craig{that uses} a transformer encoder based on BERT~\cite{BERT}; and SASRec~\cite{SASRec} -- a model which utilises decoder part of the Transformer (similar to GPT~\cite{gpt2}). \sd{For both models, we use the  versions\footnote{\sd{The code for the paper is available at  \href{https://github.com/asash/RecJPQ}{https://github.com/asash/RecJPQ}}} from \crcb{our} recent reproducibility paper~\cite{Bert4RecRepro}, which provides efficient \& effective implementations (using the \crcb{Huggingface Transformers} library~\cite{wolf_huggingfaces_2020})}. Additionally, to demonstrate that RecJPQ can be applied to other architectures, we use a GRU~\cite{choProperties2014}-based model from~\cite{PetrovRSS22} available in the same repository. \sd{This model uses the GRU4Rec~\cite{hidasiRecurrentNeuralNetworks2018} architecture, but a slightly different configuration, e.g.\ it uses LambdaRank~\cite{burges_ranknet_2010} as a loss function, which is shown to be effective~\cite{PetrovRSS22}}. 

\crcf{\textbf{Datasets.}}
 We experiment with \craig{three} datasets: (i) MovieLens-1M (denoted ML-1M)~\cite{harper_movielens_2015} -- this is a movie rating dataset that is one of the most popular benchmarks for recommender systems; (ii) Booking.com~\cite{goldenberg_bookingcom_2021} -- a multi-destination trips dataset, and (iii) Gowalla~\cite{choFriendship2011} -- a check-in dataset. Following common practice~\cite{BERT4Rec, Bert4RecRepro}, we remove users with less than \crcf{five} interactions. Table~\ref{tb:datasets} lists the salient characteristics of the datasets after preprocessing. As \craig{can be seen} from the table, the number of items in these datasets varies from relatively small (3416 in MovieLens-1M) to large (1,271,638  in Gowalla) -- this allows testing RecJPQ in different settings (RecJPQ is designed for large datasets, and we expect it \sdt{to compress the model by a larger factor} on Gowalla). 
 
\begin{table}[tb]
\caption{\looseness -1 Impact of RecJPQ with different \crcb{sub-item id} assignment strategies on model size and effectiveness  \crcf{for the} \crcb{MovieLens-1M and \crcs{Booking.com} datasets}. 
Relative Size corresponds to model checkpoint size as the percentage of the base model. $^=$,  $^+$, and $^-$ denote significance testing results compared to the base, respectively: indistinguishable ($pvalue > 0.05$, Bonferroni multi-test correction), better or worse. Bold denotes the best NDCG@10 in each column. 
}\vspace{-0.7\baselineskip}\label{tb:results_table}\
\resizebox{\linewidth}{!}{
  \begin{tabular}{l|ll|ll|ll}
\toprule
Model$\rightarrow$ & \multicolumn{2}{c|}{BERT4Rec} & \multicolumn{2}{c|}{GRU} & \multicolumn{2}{c}{SASRec} \\
Strategy$\downarrow$ & \makecell[c]{NDCG \\ @10} & \makecell[c]{Relative \\ Size} & \makecell[c]{NDCG \\ @10} & \makecell[c]{Relative \\ Size} & \makecell[c]{NDCG \\ @10} & \makecell[c]{Relative \\ Size} \\
\midrule
\multicolumn{7}{c}{ML-1M}\\ 
 \midrule
Base & \textbf{0.157} & 100.0\% & 0.072 & 100.0\% & \textbf{0.131} & 100.0\% \\
\swsdm{Hashing (Quotient-Remainder)} & 0.040$^{-}$ & 92.4\% & 0.017$^{-}$ & 61.6\% & 0.009$^{-}$ & 124.9\% \\
RecJPQ-BPR & 0.156$^{=}$ & 93.2\% & \textbf{0.076$^{=}$} & 62.5\% & 0.130$^{=}$ & 128.0\% \\
RecJPQ-Random & 0.156$^{=}$ & 93.2\% & 0.075$^{=}$ & 62.5\% & 0.125$^{=}$ & 127.6\% \\
RecJPQ-SVD & 0.154$^{=}$ & 93.2\% & 0.074$^{=}$ & 62.5\% & 0.129$^{=}$ & 127.9\% \\
\midrule
\multicolumn{7}{c}{Booking}\\ 
 \midrule
Base & 0.376 & 100.0\% & 0.209 & 100.0\% & 0.137 & 100.0\% \\
\swsdm{Hashing (Quotient-Remainder)} & 0.192$^{-}$ & 62.8\% & 0.186$^{-}$ & 27.6\% & 0.014$^{-}$ & 9.2\% \\
RecJPQ-BPR & 0.375$^{=}$ & 63.3\% & \textbf{0.334$^{+}$ }& 27.5\% & 0.242$^{+}$ & 8.7\% \\
RecJPQ-Random & 0.316$^{-}$ & 62.3\% & 0.324$^{+}$ & 27.5\% & \textbf{0.256$^{+}$} & 8.9\% \\
RecJPQ-SVD & \textbf{0.379$^{+}$} & 63.3\% & \textbf{0.334$^{+}$} & 27.6\% & 0.185$^{+}$ & 8.8\% \\\bottomrule
\end{tabular}

}
\vspace{-1.3\baselineskip}
\end{table}

\looseness -1 The datasets are also diverse regarding the number of ``long-tail items'' (defined as items with less than five interactions). While the MovieLens-1M dataset does not have long-tail items, the Booking.com dataset has 60.8\% long-tail items, and Gowalla has 75.8\% long-tail items. As discussed in Section~\ref{ssec:recjpq_is_reg}, RecJPQ acts as a model regulariser in long-tail distributions, and we expect to see the highest regularisation effect on the Gowalla dataset. 

\crcf{\textbf{Evaluation Protocol.}}
\looseness -1 Overall, our evaluation protocol follows the protocol from the \craig{recent} replicability paper~\cite{Bert4RecRepro}.  We use a leave-one-out data splitting strategy: we hold out the last sequence in each sequence in the test set. Additionally, for 1024 randomly selected users, we hold out the second last action into \craig{a} validation set, which we use for the early stopping mechanism.  We set the maximum sequence length at 200. If the sequence contains more than 200 interactions, we use 200 latest interactions. If the sequence contains less than 200 interactions, we left-pad it to ensure its length is exactly 200. \sd{To ensure that the models \swsdm{are fully converged}, following~\cite{PetrovRSS22}}, we employ an early stopping mechanism on the NDCG@10 metric: we stop training if the metric is not improved for 200 epochs.

\looseness -1 \crcf{\textbf{Metrics.}}
\crcf{Our main focus} is the trade-off between model size and model \swsdm{effectiveness. For measuring effectiveness}, following prior research~\cite{SASRec, BERT4Rec, Bert4RecRepro}, we use NDCG@10, and as the model size metric, we use the file size of the model checkpoint. Following \crcf{best practices}~\cite{canamares_target_2020, krichene_sampled_2022, dallmann_case_2021}, we measure NDCG without using negative sampling. 

\pageenlarge{3}
\crcf{\textbf{Baselines.}}
\sd{We deploy an adaptation of Quotient Remainder~\cite{shiCompositionalEmbeddingsUsing2020} \craig{as a baseline compression approach, applied to each base model} -- this \craig{parameter-free} \swsdm{hashing-based} approach encodes each item using two \swsdm{hashes}: the quotient and the remainder of the division of item id by $\left\lceil \sqrt{|I|}\right\rceil$ where $|I|$ is the catalogue size. Quotient Remainder guarantees that each item has a unique code.}

\looseness -1 We do not apply post-training embedding quantisation (e.g.\ float16), nor use other methods from Table~\ref{tb:model_compression} as baselines, as they are not suitable for our task: EODRec, LightRec, MDQE and MGQE require training full embeddings (we assume that training full embeddings is not an option for a large catalogue), and PreHash is specific for compressing user embeddings, so is not suitable for item embeddings. \sd{However, reducing the model size by decreasing the embedding dimensionality can also be seen as a simple baseline. We analyse models using different embedding sizes in \crcf{RQ3}.}%

To analyse the effect of the \crcb{sub-item id} assignment strategy on model performance/model size tradeoff, we compare the original (base) versions of BERT4Rec, SASRec and GRU with RecJPQ versions trained with Random, discrete truncated SVD and discrete BPR \crcb{sub-item id} assignment strategies. We do not train GRU and BERT4Rec on Gowalla, as these models do not use negative sampling. Training models on this dataset without negative sampling is not feasible due to the large GPU memory requirement for storing \sd{output} scores~\cite{PetrovRSS22}, \swsdm{while applying negative sampling is a substantial change to the models' training process that is outside of the scope of this paper}. In all cases, we use 512-dimensional embeddings and the code of length $m=8$ (we experiment with other embedding sizes and lengths of the code in the next section). One exception is \crcs{for the SASRec base} model on Gowalla; in this case, we use 128-dimensional item embeddings (item embeddings larger than 128 dimensions \craig{consume all available GPU memory when embedding compression techniques are not deployed).}

\vspace{-1.2\baselineskip}
\subsection{Results} \label{ssec:result}
\looseness -1 \textbf{\crcf{RQ1. Effect of \crcb{sub-item id} assignment strategy.}}
\looseness -1 Table~\ref{tb:results_table} \craig{shows the experimental} results on \swsdm{the smaller ML-1M and Booking datasets, while Table~\ref{tb:results_table_gowalla} reports results for the Gowalla dataset}. \sd{The tables compare NDCG@10 and model size of compressed variations of backbone models with the base (uncompressed) model. \craig{Significant differences compared to the corresponding base model (BERT4Rec, GRU or SASRec) are indicated.}  In general, the tables show that} RecJPQ substantially reduces the model checkpoint size in most cases. For example, the RecJPQ versions of the GRU models on the Booking dataset are approximately 27\% of the original in size. On the Gowalla dataset, compressed models are approximately 3\% of the original. Moreover, model size does not depend on the \crcb{sub-item id} assignment strategy. Indeed, \crcb{sub-item id} assignments only influence the values of the model parameters but not the number of parameters. Moreover, Quotient Remainder models have approximately the same compression level as RecJPQ models. We speculate that after compression, the model checkpoint size is dominated by other model parameters (e.g., attention matrices). In our configuration, the \sd{\crcb{sub-item} embeddings tensor} only requires a few megabytes of memory (see Table~\ref{tb:centroids_memory}). In contrast, the full model checkpoint of a compressed model is typically tens of megabytes (e.g., 92.8MB for SASRec using RecJPQ-BPR trained on Gowalla). 

\pageenlarge{3}
\looseness -1 
RecJPQ only increased the model size for SASRec on \crcs{MovieLens-1M,} due to the dataset's small item count: \crcf{on this dataset, }the overhead of storing \crcb{sub-item} embeddings and the codebook  \crcf{is larger compared to} the benefit of compressing the embeddings table. Using RecJPQ with smaller embeddings might reduce the model size without affecting \crcf{effectiveness} on this dataset (see \crcf{also RQ3}).

On the other hand, we \craig{observe} from \swsdm{Table~\ref{tb:results_table}} and \swsdm{Table~\ref{tb:results_table_gowalla}} that the choice of the \crcs{best assignment} strategy depends on both the model and the dataset. For example, on MovieLens-1M, the choice of the strategy is \craig{not} important, and in all cases, RecJPQ versions of the models are statistically indistinguishable \crcs{from all corresponding base models}. 
On the larger Booking dataset, the choice of the best strategy is model-dependent. For BERT4Rec, the best results are achieved with BPR (NDCG@10 0.375, statistically indistinguishable from the base) and SVD (NDCG@10 0.379, +0.97\%, significant).  At the same time, the Random strategy significantly underperforms the base configuration (NDCG@10 0.316, -15.98\%) -- this shows that in some cases, assigning similar codes to similar items is indeed important.  However, in 2 cases,  Random performs statistically significantly better than SVD and BPR.  For example, Random assignments perform best on Gowalla with the SASRec base (a significant improvement of +57\%  over the base). SVD assignments also moderately improve the result in this case (+10\%, significant). At the same time, BPR decreases the quality by a large margin \sd{on Gowalla dataset} (-70\%)\footnote{\sd{The percentages for the Gowalla dataset seem large because this dataset is difficult: it has the largest number of items and the largest proportion of long-tail items.}}.  We explain the success of the Random strategy on the Gowalla dataset as giving a larger regularisation effect (random assignments make the learning task harder, so the model has fewer chances to overfit).
\crc{Overall, from Tables 4 and 5, we \crcb{conclude} that there is no "one size fits all" choice of the \crcb{sub-item id} assignment strategy, and it depends on both dataset and model salient characteristics; the exact best combination of model/strategy may depend on regularisation requirements (as we observe in Gowalla), the prevalence of strong sequential patterns (as in Booking, see~\cite{PetrovRSS22} for details) and so on.} This suggests that the \crcb{sub-item id} assignment strategy \sd{could} be treated as a hyperparameter and tuned for each model/dataset combination. \sd{However, by default,} we recommend using RecJPQ with \crcb{the} SVD strategy -- in all cases, \crcb{this} achieves significantly better (on the Booking and Gowalla datasets) or statistically indistinguishable (on the MovieLens-1M dataset) results compared to the base model. We also note that RecJPQ with the SVD strategy is always better than the Quotient Remainder baseline (Quotient Reminder is always significantly worse than the base \crcb{model}, whereas RecJPQ is better or indistinguishable). \crc{The question of whether or not it is possible to select the best model/strategy combination without doing an exhaustive hyperparameter search is an interesting research direction, which we leave for future work.}

\begin{table}[t]
  \caption{ Impact of RecJPQ with different \crcb{sub-item id} assignment
  strategies on SASRec model size and effectiveness on the large-scale Gowalla dataset. 
  Notations follow Table~\ref{tb:results_table}.}\label{tb:results_table_gowalla}
  \vspace{-0.7\baselineskip}
  \resizebox{0.6\linewidth}{!}{
    \swsdm{
     \begin{tabular}{l|ll}
\toprule
Strategy & \makecell[c]{NDCG @10} & \makecell[c]{Relative Size} \\
\midrule
Base & 0.110 & 100.0\% \\
\swsdm{Hashing (Quotient-Remainder}) & 0.081$^{-}$ & 2.8\% \\
RecJPQ-BPR & 0.033$^{-}$ & 2.8\% \\
RecJPQ-Random & \textbf{0.173$^{+}$} & 2.9\% \\
RecJPQ-SVD & 0.122$^{+}$ & 2.9\% \\
\bottomrule
\end{tabular}

     }
  }
  \vspace{-1\baselineskip}
\end{table}

\begin{figure}
    \vspace{-0.5\baselineskip}
    \subfloat[MovieLens-1M]{
      \includegraphics[width=0.42\textwidth]{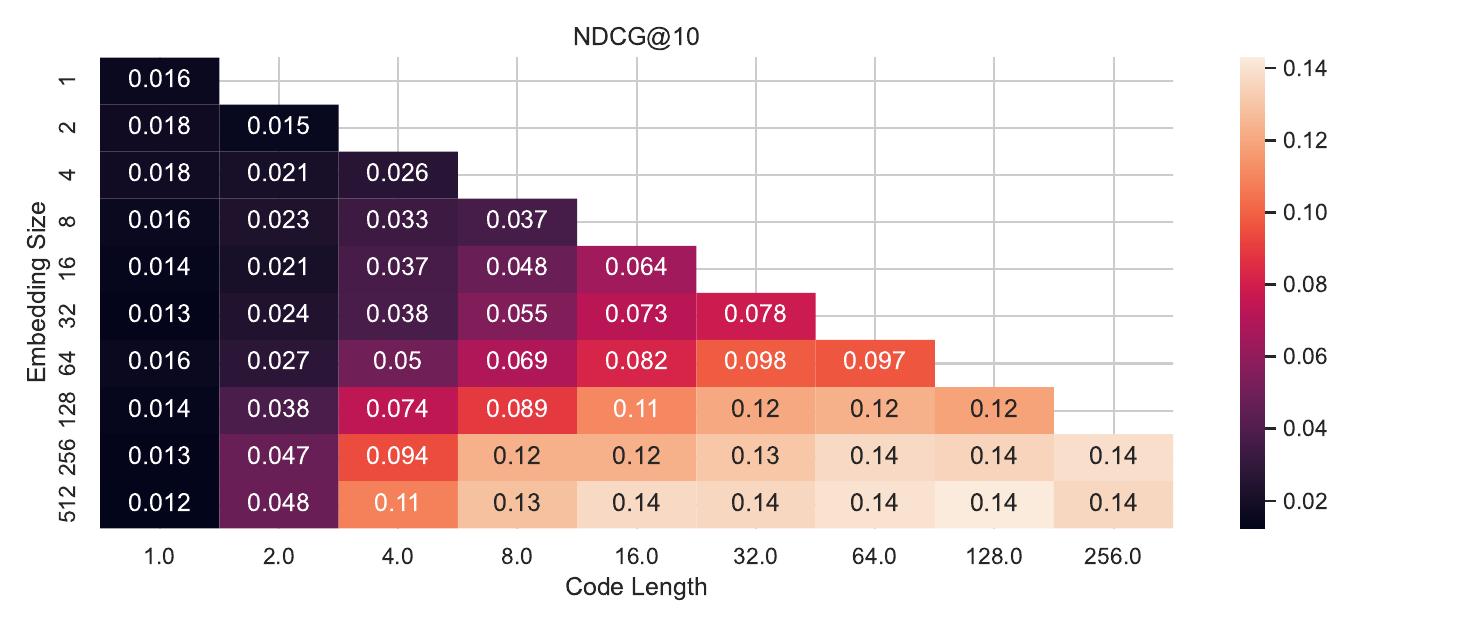}
      \label{subfig:grid_search:ml}
    }
    \\ \vspace{-1\baselineskip}
    \subfloat[Gowalla]{
      \includegraphics[width=0.42\textwidth]{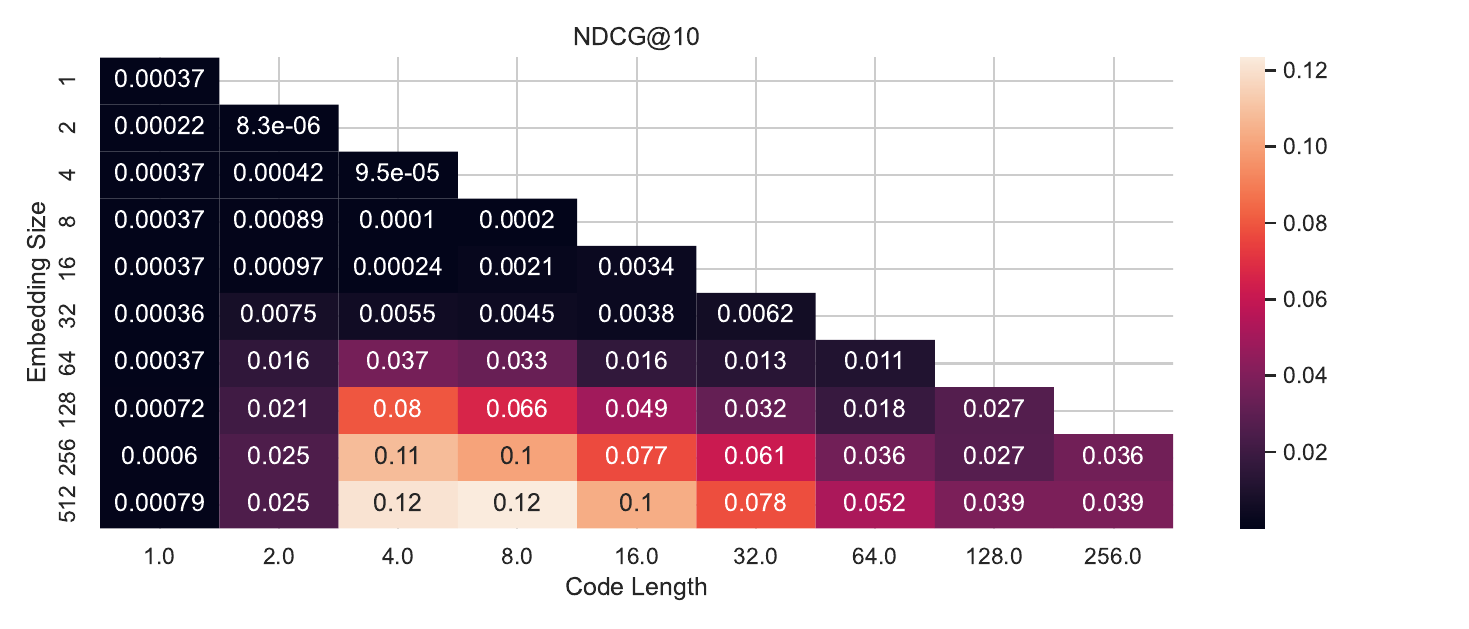}
      \label{subfig:grid_search:gowalla}
    }
    \vspace{-1\baselineskip}
    \caption{RecJPQ performance \crcb{for SASRec} while varying embedding size~$d$ and the number of \crcb{sub-item ids} per item $m$.}
\label{fig:grid_search}
     \vspace{-1\baselineskip}
\end{figure}

\pageenlarge{3}
\looseness -1 \swsdm{It is also worth mentioning} that we do not observe a degradation of the training efficiency when training the RecJPQ versions of the models. Indeed, while there are some fluctuations in the training time the model requires to converge, the magnitude of the required training time remains the same: for example, training of the base version of BERT4Rec requires 18.8 hours on Booking.com, and the RecJPQ-SVD version of BERT4Rec requires 16.1 hours. The training time of the SVD model (used for initial \crcb{sub-item id} assignments) in the same case is negligible compared to the training of the main model (it takes approximately a minute). \swsdm{Inference time is also unaffected (e.g.\ on Gowalla, full evaluation across the 86k users requires 10 minutes for both "base" and RecJPQ versions of the models).} %

\looseness -1 In summary, in answer to RQ1, we conclude that RecJPQ achieves large model compression levels. \sd{\crcf{The compression} is particularly impressive on} datasets with large catalogues (like Gowalla). The compression does not depend on the \crcb{sub-item id} assignment strategy. However, the \crcb{sub-item id} assignment strategy greatly affects the model performance. The effect is model- and dataset-dependent, so the strategy should be treated as a hyperparameter. However, the SVD is a safe choice, as it always provides results that are comparable (i.e statistically indistinguishable) or better than the base model.

\looseness -1 \textbf{\crcf{RQ2. Effects of code length $m$ and the embedding
size on model performance.}} 
To answer \swsdm{RQ2}, we perform a grid search over embedding size and code length on the MovieLens-1M and Gowalla datasets. We use SASRec as the backbone (the only model that can be easily trained on Gowalla) and apply the SVD \crcb{sub-item id} assignment strategy. We select the embedding size $d$ from \{$2^0, 2^1, ..., 2^9$\} and code length $m$ from \{$2^0, 2^1, ..., 2^8$\}. \swsdm{Note that $m\leq d$, as RecJPQ splits each} embedding of size $d$ into $m$ sub-embeddings. 

\pageenlarge{3}
\looseness -1 Figure~\ref{fig:grid_search} illustrates the results of the grid search. \sd{The figure shows the NDCG@10 of the SASRec-RecJPQ model for each combination of code length (x-axis) and embedding size (y-axis), in the form of a heatmap for both datasets}. As we can see from the figure, a larger embedding size generally positively affects the model performance. This result echoes similar findings of a recent reproducibility paper~\cite{rendles_revisiting_2022}; however, interestingly, in the RecJPQ case, increasing
embedding dimensionality does not change the amount of information we store per each item, as the length of the code defines it rather
than the embedding size. Instead, it increases model capacity, increasing the 
amount of information that can be stored in each \crcb{sub-item embedding}, allowing %
to account for more item characteristics.  For example, on Gowalla, the largest embedding we can train using base SASRec is \crcb{128 dimensions}, while with RecJPQ, we can train the model even with 512-dimensional embeddings. 

On the other hand, larger code lengths are not always helpful. As we discussed in Section~\ref{ssec:recjpq_is_reg}, RecJPQ forces the model to share parts of embeddings with other items and, therefore, acts as a regularisation mechanism. A shorter code length forces items to share more information, and therefore, it causes a stronger regularisation effect. As we can see, on the less sparse MovieLens-1M -- where all items have more than five interactions -- regularisation is not an issue, and longer codes are beneficial. For example, the best result is achieved with 512-dimensional embeddings and a code of length 128 (NDCG@10 0.14). In contrast, for Gowalla, where most items are long-tail items with less than five interactions (hence the embeddings of these items should be regularised), the best NDCG is achieved with the code of length 8 (NDCG@10 0.12). The fact that the model can perform better with shorter codes confirms that RecJPQ can \crcs{behave} as a regularisation technique. 

\pageenlarge{3}
In short, in answer to RQ2, we conclude that larger embeddings are generally beneficial for model performance. However, the sparser Gowalla dataset benefits from shorter code lengths, due to the regularisation effect of parameter sharing brought by RecJPQ.

\begin{figure}
\centering
 \vspace{-1.0 \baselineskip}
  \subfloat[MovieLens-1M]{
    \hspace{-0.5cm}\includegraphics[width=0.26\textwidth]{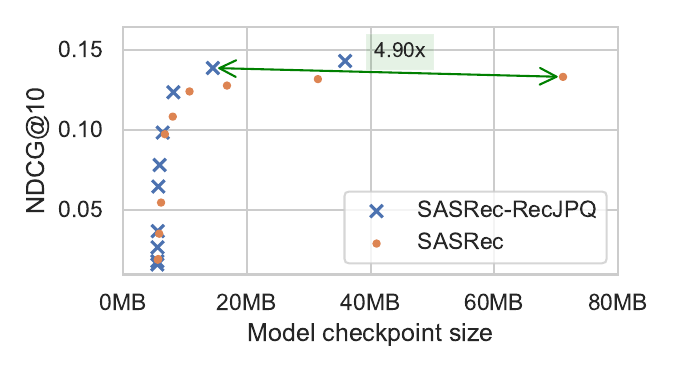}
    \vspace{-0.5\baselineskip}
  }
  \subfloat[Gowalla]{
    \includegraphics[width=0.26\textwidth]{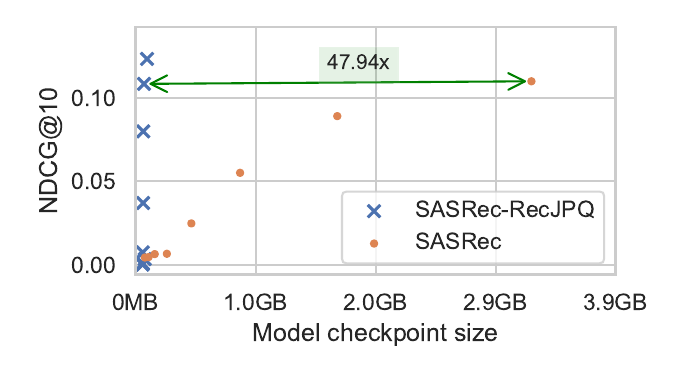}
    \vspace{-0.5\baselineskip}
  }
  \vspace{-0.5\baselineskip}
  \caption{\crcf{NCDG/Size} tradeoff for SASRec and SASRec-RecJPQ.}
  \label{fig:ml_1m_ndcg}
  \vspace{-1 \baselineskip}
\end{figure}

\textbf{\crcf{RQ3. Size-Performance tradeoff.}}
To address our last research question, we analyse the trade-off between model checkpoint size and NDCG@10 achieved by the model when trained with different embedding sizes. We select the embedding size from \{1, 2, 4, 8, 16, 32, 64, 128, 256, 512\} and train the original versions of SASRec and SASRec-RecJPQ with the SVD strategy on \craig{the} MovieLens-1M and Gowalla datasets. For RecJPQ, we select code length $m$ \crcb{that is} optimal for the dataset/embedding size pair (according to \crcb{the} grid search from Figure~\ref{fig:grid_search}). For the original SASRec on Gowalla, we only train up to the \sd{embedding} size of 128 \crcf{due to GPU memory limit}.

\looseness -1 Figure~\ref{fig:ml_1m_ndcg} illustrates the tradeoff between model checkpoint size and NDCG@10 for both SASRec and SASRec-RecJPQ on the two datasets. Each point on the figure corresponds to one embedding size. As can be seen from the table,  a larger model size (corresponding to larger embeddings) leads to better performance for both SASRec and SASRec-RecJPQ (this echoes findings in the previous research question). However, SASRec-RecJPQ's performance grows much faster \craig{with increasing model size} than \craig{observed for} vanilla SASRec. For example, the largest vanilla SASRec model achieves roughly the same performance as the \swsdm{4.9$\times$ smaller SASRec-RecJPQ version of the model} (\sd{71MB vs.\ 15 MB}). This effect is even more prominent in Gowalla, where the number of items is \swsdm{larger: the largest SASRec model achieves roughly the same performance as the 47.94$\times$ smaller SASRec-RecJPQ model (3.2GB vs.\ 69MB)}. 

\looseness -1 Overall, in answer to RQ3, we conclude that while larger models benefit model performance, RecJPQ improves this tradeoff by a large margin \swsdm{(i.e.\ to achieve the same performance, RecJPQ requires much fewer parameters than the original model)}. This effect is more markedly pronounced for the larger Gowalla dataset.

\vspace{-0.5\baselineskip}
\section{Conclusions}\label{sec:conclusion}
\looseness -1 In this paper, we discussed the challenge of training sequential recommender systems with large datasets, primarily due to the large item embedding tensor. Existing \crcb{embedding} compression methods have limitations, leading to \crcs{our} proposed method, RecJPQ. Our evaluation of  RecJPQ on three datasets resulted in significant model size reduction, e.g., \swsdm{47.94$\times$} compression of the SASRec model on the Gowalla dataset. Additionally, RecJPQ serves as a model regulariser, improving the model's quality, with SASRec-RecJPQ using SVD strategy outperforming the original SASRec model (+35\% NDCG@10 on Booking, +10\% on Gowalla). \crc{There are a number of future \crcs{work} directions, which can be based on this \crcs{paper}, e.g., an analysis of the RecJPQ effect on training/inference time and analysing \crcs{the} theoretical grounds for the observed regularisation effects. }

\FloatBarrier
\balance
\bibliographystyle{ACM-Reference-Format}
\bibliography{references}
\end{document}